\documentstyle[preprint,aps,eqsecnum,tighten]{revtex}
\preprint{\vbox{\baselineskip=12pt
\rightline{CGPG-96/3-3}
\rightline{gr-qc/9603030}}}

\def\be{\nopagebreak[3]\begin{equation}}
\def\ee{\end{equation}}
\def\ba{\nopagebreak[3]\begin{eqnarray}}
\def\ea{\end{eqnarray}}
\def\nl{\nonumber \\}
\def\ni{\noindent}

\def\a{\alpha}
\def\b{\beta}
\def\c{\gamma}
\def\d{\delta}
\def\e{\eta}
\def\ve{\varepsilon}
\def\f{\phi}

\def\m{\mu}
\def\n{\nu}

\def\th{\theta}
\def\r{\rho}
\def\s{\sigma}

\def\w{\wedge}

\newcommand{\al}{\mbox{$\hskip -.12cm \ _{\alpha}\hskip -.02cm$}}

\begin{document}
\title{
Topological Lattice Gravity\\
Using Self-Dual Variables 
}
\author {
Jos\'e A. Zapata\thanks{zapata@phys.psu.edu}
}
\address{
Center for Gravitational Physics and Geometry \\
Department of Physics, The Pennsylvania State University \\
104 Davey Laboratory, University Park, PA 16802
}
\maketitle
\begin{abstract}
Topological gravity is the reduction of general relativity to 
flat space-times. A lattice model describing topological gravity 
is developed starting from a Hamiltonian lattice version of 
$B\w F$ theory. The extra 
symmetries not present in gravity that kill the local degrees of 
freedom in $B\wedge F$ theory are removed. 
The remaining symmetries preserve the geometrical character of the 
lattice. 
Using self-dual variables, the conditions that guarantee the 
geometricity of the lattice become reality conditions. 
The local part of the remaining symmetry generators, that respect 
the geometricity-reality conditions, has the form of Ashtekar's 
constraints for GR. 
Only after constraining the initial data to flat lattices and 
considering the non-local (plus local) part of the constraints 
does the algebra of the symmetry generators close. 
A strategy to extend the model for non-flat connections and 
quantization are discussed. 
\end{abstract}
\pacs{PACS number(s): }
\clearpage 

\section{Introduction}\label{Introduction}

 Regularization appears as an inescapable step in the quantization 
of most interacting field theories. Ashtekar's non-perturbative 
quantization program for gravity \cite{ash1} is no exception; suitable 
regularization is needed to construct 
geometric operators and the constraints 
selecting the physical states. 
The standard expressions for the geometric operators to measure 
areas and volumes 
\cite{rovelli-smolinAREA,ALMMT}, and the scalar constraint 
\cite{rovelli-smolinLOOP,pullinLOOP} were derived 
through point-splitting regularization; on the other 
hand, the pioneering work of Renteln and Smolin \cite{renteln-smolin} 
on lattice regularization remained quite distant from the main stream. 

Paradoxically, ideas from lattice gauge theory have played a dominant
role in the connectiondynamic approach 
to quantum gravity. Lattice gauge theory is used not to
replace the continuum by a lattice as a regularization step, but 
to rigorously 
define the quantum configuration space of connectiondynamics 
$\overline{{\cal A}/{\cal G}}$. 
The spaces of connections (modulo gauge transformations) of all
the graphs embedded in the manifold are linked by a projective 
structure to define  
$\overline{{\cal A}/{\cal G}}$; 
for a review of the field see \cite{ALMMT,baez}. 

Apart from the mentioned work intensive research in other approaches 
to quantum gravity and related areas like the 
dynamical triangulation approach and the Turaev-Viro model 
provide strong motivations to further develop 
lattice regularization within Ashtekar's quantization program. 

The recent work by Loll \cite{loll} and Immirzi \cite{immirzi}, 
and the no so recent 
by Waelbroeck \cite{waelbroeck} and Katsymovsky \cite{katsymovsky}  
share some of 
the mentioned motivations; this research on lattice 
regularizations for Ashtekar's 
quantization program has a geometrodynamic relative: 
the original Regge calculus \cite{regge}. 
One must recall that all the different attempts to 
construct an initial value formalism of Regge calculus as a 
fundamental theory have failed; their constraints are first-class only 
in the continuum limit. 

This article presents a classical connectiondynamic model of 
$3+1$ Regge calculus. From this classical toy model one can get 
hints to solve some problems that may occur in future 
lattice regularization of the constraints. 
The model describes flat space-times as the evolution 
of a three-dimensional simplicial lattice. 
It is based on a $SO(3,1)$ lattice gauge theory where 
every cell has four neighbors. In addition, the variables of the 
theory are required to satisfy some 
``geometricity conditions''. Once the geometricity 
conditions are fulfilled, the variables of the lattice gauge theory 
specify the geometry of 
a three-dimensional piecewise linear space that generates
space-time as it evolves. In terms of self-dual variables 
the model for Regge calculus acquires an Ashtekar-like description. 
After changing to self-dual variables the 
geometricity conditions take the form of the reality conditions 
of Ashtekar's GR. Spatial and 
time-like translations that preserve the geometricity conditions 
are generated by constraints; the local part of these 
symmetry generators 
resembles the constraints of general relativity written in terms of 
self-dual variables. The phase space variables of the model and the 
geometricity conditions of the lattice are closely related 
to the ones given by 
Immirzi in \cite{immirzi}; see also \cite{katsymovsky}. 

The origin of this model is a $2+1$ lattice theory formulated by 
Waelbroeck \cite{waelbroeck}, but a closer relative is the extension 
of the $2+1$ theory: a lattice $B\wedge F$ theory in $3+1$ dimensions 
(Waelbroeck and Zapata \cite{tlg}). In the lattice $B\w F$ (LBF) case 
the geometricity conditions fixed a ``geometrical gauge'', 
a precise statement of 
what the geometricity conditions mean in the LBF case is stated in 
section~\ref{constraints}. Once in the geometrical gauge the $B\wedge F$ 
constraints take the form of the 4d-translation 
generators of the vertices of 
the lattice; these constraints together with the geometricity conditions 
render the lattice's connection flat, making an extension of the theory to 
lattices with curvature 
not viable. Another manifestation of the same issue is the fact that 
the lattice $B\w F$ theory has no degrees of freedom 
associated with the lattice vertices or cells. The only possible 
degrees of freedom are topological \cite{waelbroeckBF}. 

In this article I introduce two results that help in the construction of 
a bridge between lattice $B\w F$ (LBF) and lattice gravity.
Firstly, self-dual variables in the model 
allow the geometricity conditions, that in LBF were ordinary second-class 
constraints, to be treated as reality conditions. 
Secondly, the model's symmetry group is effectively smaller than 
that of LBF and the symmetry generators do not restrict the 
lattice to be flat making viable 
an extension of the model to a theory with local degrees of freedom.  
A vector and a scalar constraint 
(per lattice cell) replace the four dimensional covariant generators of 
vertex translations of LBF. Remarkably, the local 
part of these new constraints has the form of the 
Ashtekar's constraints which, as explained in sec.~\ref{constraints}, 
implies that the continuum limit of the model is Ashtekar's
formulation of GR. 
An important aspect of this article is the notation. 
The mentioned relation between geometricity and reality conditions and 
the indication that the local part of the constraints are Ashtekar-like, 
are highly clarified after the introduction of an ``affine'' notation 
for the lattice \cite{sorkin+}. The affine notation indicates 
``space directions'' in a manner natural for the discreteness of the 
lattice, while resembling the notation used in the continuum. 
The ``affine'' notation simplifies the difficult task of 
translating physically meaningful expressions from the continuum to 
the lattice \cite{brewin}, thus, providing a useful tool for $3+1$ 
Regge calculus. 

Once the relation between the model and canonical continuum gravity 
has been realized, it is natural to ask about quantization. 
One would like to adapt the original version of Ashtekar's quantization 
program \cite{ash1} to lattice gravity. 
To this end one would select the physical Hilbert space 
following Dirac's prescription and then fix the inner product 
to make former real quantities Hermitian operators. 
However, in a study on the quantization of Regge calculus 
\cite{immirziQUAN} Immirzi pointed out that 
the plan of choosing 
the inner product according to the quantum reality conditions fails 
when following the conventions derived from 
canonical gravity. The plan's failure is 
nothing but another manifestation of the parallelism 
between the particular lattice approach to quantum gravity followed in 
this article and the approach of Ashtekar and collaborators 
for continuum gravity. In the context of continuum gravity 
Thiemann \cite{thiemannHALL,thiemannCOMPL} introduced a generalized Wick
transform to implement the quantum reality conditions. Some implications of
importing Thiemann's strategy to lattice gravity are discussed in the
concluding section. 

The organization of the article is the following. Section~\ref{framework} 
describes the 
framework of the lattice theory. It presents the affine notation, a review 
of the lattice $B\wedge F$ theory, and introduces self-dual variables 
for the lattice. 
Section~\ref{geo-reali} contains a derivation of the geometricity conditions 
and its expression as reality conditions. 
In section~\ref{constraints}, the constraints of the model 
are introduced, as well as the 
induced symmetries and their algebra. The possibility of extending this 
model to space-times with curvature is thoroughly discussed 
in the concluding section. 

\section{Framework}\label{framework}
\subsection{Affine Notation}\label{affine}

 A convenient tool for translating expressions from the 
continuum to a lattice framework is the affine notation \cite{sorkin+}. 
In a n-dimensional simplicial 
lattice $\Sigma$, the analog of its 
tangent bundle is a collection of 
n-dimensional vector spaces (one attached to every cell of the 
lattice). In each of these vector spaces, the $n+1$ intrinsically defined 
bivectors related to the boundary faces of the simplex and the natural 
volume element select $n+1$ one-forms. These intrinsically defined 
one-forms $(\omega ^j)_a$ ($j=1,...,n+1$ label the one-forms, 
and $a$ is an abstract Minkowski index) 
can play the role of an affine basis. 
For any one-form $\sigma _a$ 

\ba 
\sigma _a= (\omega ^j)_a \sigma_j 
\ea 
\ni
where 
\ba 
{\sum _j} (\omega ^j)_a = 0 \quad ,
\quad {\sum _j} \sigma_j = 0 \quad , 
\ea 

\ni
the first condition holds because the lattice is formed by closed 
cells and the second condition guarantees the uniqueness of 
the affine components $\sigma_j$. 
One can construct a dual basis of vectors $(e_j)^a$ 
at each cell from the condition 

\ba 
(\omega ^j)_a (e_k)^a = \hat{\delta}^j_{\ k}
\ea 
\ni
where 
\ba
\hat{\delta}^j_{\ k} = \delta ^j_{\ k} - \hat{n}^j \hat{n}_k \quad , 
\quad \hat{n}^j = \hat{n}_j = 
{1\over \sqrt{n+1}}  
\ea

\ni
in particular, for a three dimensional space 
\be 
\hat{n}^j = \hat{n}_j = {1\over 2}
\quad , \quad \hat{\delta}^j_{\ j} = 
{3\over 4} \quad , \quad \hat{\delta}^j_{\ k\neq j} 
= -{1\over 4}. 
\ee

\ni
One can see that the projector $\hat{\delta}$ satisfies 
$\hat{\delta}^j_{\ k}\hat{\delta}^k_{\ l} = \hat{\delta}^j_{\ l} = $ 
and $\hat{\delta}^j_{\ j} = n$. A basis for tensors of higher order can 
be constructed directly from these two. 
If one labels the lattice 
cells by Greek letters $\a , \b , ...$ in such a way that cell 
$\a$ has as neighbors $\b , \c , ...$, then the directions 
of the affine basis of the vector space of cell 
$\a$ can be labeled by its neighbors $j=\b , \c , ...$. 
 
To any p-form in a manifold, with components $\sigma (x)_{j_1,...,j_p}$, 
one assigns 
a lattice counterpart $\sigma{\scriptstyle (\a)}_{j_1,...,j_p}$ 
(where ${\scriptstyle (\a)}$ plays the role of the base point
$(x)$). The lattice p-form can be regarded as a p-cochain, that is, a 
linear function that assigns real numbers to the p-chains 
of the lattice (e.g. 
a 2-cochain assigns numbers to the faces of the lattice). 
I should emphasize that 
$\sigma{\scriptstyle (\a)}_{j=\b}\equiv \sigma{\scriptstyle (\a)}_{\b}$ 
should be regarded as the lattice counterpart of e.g. 
$\sigma (x)_{\underline{j}=1}\equiv \sigma (x)_1$; therefore, the
Einstein summation 
convention should not be applied for Greek indices appearing in the right. 
Using the affine notation, one can easily 
find the discrete counterpart of the $B\wedge F$ action.

\subsection{Lattice ${\bf B\wedge F}$ Theory} 

 This subsection mirrors part of \cite{tlg}; however, 
it is reviewed here using the affine notation for the convenience of the 
reader (the notation and conventions of this paper 
follow Peld\'an in \cite{peldan}, where he gives a 
formulation of continuum GR using $SO(3,1)$ as internal group). 

The $B\wedge F$ theory, as Horowitz formulated it \cite{horowitz}, 
starts from a modified Palatini action 

\be
S_{\small B\wedge F} = \int _{\cal M} B_A \wedge F^A 
\label{3}
\ee

\ni
where the internal group is $SO(3,1)$ and the indices $A$ can be written 
as $A=[ab]$, $a,b=0,1,2,3$. In this modified Palatini action the 
constraint on the $so(3,1)$ valued two-form 

\be
B_A := B_{[ab]}= \varepsilon_{abcd}\  e^c \wedge e^d 
\label{4}
\ee
\ni
has been dropped, making gravity and $B\wedge F$ theory different theories.

A space-like discrete version of $B\wedge F$ theory can be formulated 
from the discrete counterpart of the $3+1$ split of the $B\wedge F$ 
action \cite{horowitz}:

\ba 
S[B,A]_{\small B\wedge F}= 3 \int dx^0 \int _{\Sigma }
(\dot A ^A_{[i} B_{jk]A}-
F^A_{[ij}B_{k]0A} 
+A ^A_0 D_{[i} B_{jk]A})dx^idx^jdx^k
\label{}
\quad .
\ea

\ni
The lattice counterpart of the last expression should be 
considered as the starting point of this lattice formalism \cite{tlg} 

\be
S[E,M]=3\int dx^0 {\sum _{\a}}(4{\bf E}{\scriptstyle (\a)}^j\cdot
{\bf \dot A}{\scriptstyle (\a)}_j  
-{\bf P}{\scriptstyle (\a)}_{jk}\cdot {\bf E}{\scriptstyle (\a)}^{jk} 
+{\bf A}{\scriptstyle (\a)}_0\cdot {\bf J}{\scriptstyle (\a)} )
\label{6}
\ee

\ni
where $E{\scriptstyle (\a)}^j_A:={1\over 8}B{\scriptstyle (\a)}_{klA}
\hat{\varepsilon }^{jkl}$, 
$\hat{\varepsilon }^{jkl}:= \varepsilon ^{jklm}\hat{n}_m$, 
and the action is a functional of the variables 
$E{\scriptstyle (\a)}^{\b }_A , M{\scriptstyle (\a)}_{\b A}^{\ \ B}$
attached to every face $(\a , \b )$ 
of the lattice. 
In the action, the curvature form was replaced by 
$P{\scriptstyle (\a)}_{jk}^A =F{\scriptstyle (\a)}_{jk}^A + O(F^3)$, where 
$P{\scriptstyle (\a)}_{j=\b k=\c}^A :={1\over 4}f^{AB}_{\ \ C}
W{\scriptstyle (\a)}_{\b \c B}^{\ \ \ C}$, 
and $W{\scriptstyle (\a)}_{\b \c B}^{\ \ \ C}:=
(M{\scriptstyle (\a)}_{\b}M{\scriptstyle (\b)}_{\m} \ldots M(\n)_{\c}
M{\scriptstyle (\c)}_{\a})_B^{\ C} $ is the 
holonomy around the lattice link of cell $\a$ where faces $\b$ and 
$\c$ intersect. 
The three-form of the Gauss law term in the continuum action is 
replaced by the integral of ${\bf B}$ over the boundary of a 
lattice cell ${\bf J}{\scriptstyle (\a)}$. In the $3+1$ split 
$A{\scriptstyle (\a)}_0^A$, and $E{\scriptstyle
(\a)}^{jk}_A:=B{\scriptstyle (\a)}_{l0A}\hat{\varepsilon }^{jkl}$ 
are Lagrange multipliers. 

In a discrete scenario, the role of a connection is better played by 
matrices that define parallel transport along non-infinitesimal paths. 
Thus, the connection $A{\scriptstyle (\a)}_j^C$
that appears in the Lagrangian is regarded as a 
secondary quantity defined in terms of the matrix that parallel 
transports to the reference frame at cell $\a$, from its neighbor in 
direction $(j)$, by 

\ba 
\exp(A{\scriptstyle (\a)}_j^C f_{CA}^{\ \ B}):= M{\scriptstyle
(\a)}_{jA}^{\ \ B}
\ea 
\ni
In the adjoint representation the structure constants of 
$so(3,1)$ and the generators of the group are related by 
$(T_A)_B^{\ C}=f_{AB}^{\ \ C}
=f_{[ab][cd]}^{\ \ \ \ \ \ [ef]}= -  
\delta ^{\ \ \ \ r}_{[ab] \ s}
\delta ^{\ \ \ \ s}_{[cd]\ t}\delta ^{[ef]t}_{\ \ \ \ \ r}$ 
and the Lie algebra indices are raised and lowered with the Cartan 
metric $g_{AB}=-{1\over 4}f_{AC}^{\ \ D}f_{BD}^{\ \ C}$.

To write the Lagrangian explicitly in terms of the 
parallel transport matrices, one manipulates formally the kinetic term%
\footnote{
Equation (\ref{8}) is strictly correct only for an Abelian 
group, since it neglects the ordering ambiguity of the two matrices. 
However, there are only two ways to write the 
kinetic term (\ref{L}) considering that indices can be contracted only 
if they live in the same frame. One shows the equivalence between the 
other possibility and (\ref{L}) integrating by parts.
}
\cite{waelbroeck} 

\ba
{\bf f A} &=& \ln {\bf M}\\
\dot {\bf A} &=&{1\over 4} {\bf f M}^{-1}\dot {\bf M} \label{8} \\
L&=&{\sum _{\a}}(4{\bf E}{\scriptstyle (\a)}^j\cdot  \dot {\bf
A}{\scriptstyle (\a)}_j 
-{\bf P}{\scriptstyle (\a)}_{jk}\cdot {\bf E}{\scriptstyle (\a)}^{jk} 
+{\bf A}{\scriptstyle (\a)}_0\cdot {\bf J}{\scriptstyle (\a)} )\\ 
 &=&{\sum _{\a \b}} E{\scriptstyle (\a)}^{\b}_A  f_B^{\ CA}
 M{\scriptstyle (\b)}_{\a D}^{\ \ B} 
 {\dot M}{\scriptstyle (\a)}_{\b C}^{\ \ D} 
 +{\sum _{\a}} ( -{\bf P}{\scriptstyle
(\a)}_{jk}\cdot {\bf E}{\scriptstyle (\a)}^{jk} 
+{\bf A}{\scriptstyle (\a)}_0\cdot {\bf J}{\scriptstyle (\a)} ) 
\label{L}
\ea

\ni
where in the $(\a \b )$ sum there is a term for $(\a ,\b )$ 
and a term for $(\b ,\a )$ if cells $\a$ , $\b$ share a face. 
Notice that for this Lagrangian (\ref{L}) the variables 
${\bf E}{\scriptstyle (\a)}^{\b} , {\bf E}{\scriptstyle (\b)}^{\a} ,
{\bf M}{\scriptstyle (\a)}_{\b}$, 
and $ {\bf M}{\scriptstyle (\b)}_{\a}$ 
are all independent. To relate these variables to a lattice, one has 
to impose the relations 
\ba
&&E{\scriptstyle (\a)}^{\b}_A=- M{\scriptstyle (\a)}_{\b A}^{\ \ B}
E{\scriptstyle (\b)}^{\a}_B 
\label{anti}\\
&&M{\scriptstyle (\a)}_{\b A}^{\ \ C}M{\scriptstyle (\b)}_{\a C}^{\ \
B}= \delta _A^{\ B} 
\label{inv}\\
&&M{\scriptstyle (\a)}_{\b A}^{\ \ C}M{\scriptstyle (\a)}_{\b \ C}^{\
B} = \delta _A ^{\ B} 
\label{orto}
\ea

\ni
which form a second-class set with the momentum constraints coming 
from the action. Through the Dirac procedure, one gets 
\cite{waelbroeck} the result 
first derived in the context of lattice gravity 
by Renteln and Smolin \cite{renteln-smolin} 

\ba 
\{E{\scriptstyle (\a)}^{\b}_A , E{\scriptstyle (\a)}^{\b}_B \}
&=& f_{AB}^{\ \ D}E{\scriptstyle (\a)}^{\b}_D \label{EE}\\
\{E{\scriptstyle (\a)}^{\b}_A , M{\scriptstyle (\a)}_{\b B}^{\ \ C} \}
&=& f_{AB}^{\ \ D}M{\scriptstyle (\a)}_{\b D}^{\ \ C} 
\label{EM}\\
\{E{\scriptstyle (\a)}^{\b}_A , M{\scriptstyle (\b)}_{\a B}^{\ \ C} \}
&=& f_{A\ D}^{\ C}M{\scriptstyle (\b)}_{\a B}^{\ \ D} \quad .
\label{EM^-1}
\ea

\ni
Now ${\bf M}{\scriptstyle (\a)}_j$ can be considered a 
parallel transport matrix and ${\bf E}{\scriptstyle (\a)}^{j=\b}$ a 
variable related to the boundary between cells $\a , \b$, 
because the relations (\ref{anti})-(\ref{orto}) are identities 
for the Poisson brackets (\ref{EE})-(\ref{EM^-1}). 

>From the Lagrangian (\ref{}), one also obtains the Gauss law and the 
flatness constraints 

\ba
&&J{\scriptstyle (\a)}_A ={\sum _j}E{\scriptstyle (\a)}^j_A \approx 0 
\label{cerra}
\\&&
P{\scriptstyle (\a)}_{jk}^A ={1\over 4} f^{AB}_{\ \ C} 
W{\scriptstyle (\a)}_{jkB}^{\ \ \ C} =
{1\over 4} f^{AB}_{\ \ C}
({\bf M}{\scriptstyle (\a)}_j{\bf M}{\scriptstyle (\b)}_l\ldots {\bf
M}{\scriptstyle (n)}_i)_B^{\ C} \approx 0 \quad ,
\label{plani}
\ea

\ni
If the geometricity conditions presented in the next section are 
satisfied, the previous conditions can be interpreted as the requirements 
that the lattice cells close and that the parallel transport around a 
lattice link is the identity map. 
The Gauss law constraint generates gauge transformations 

\ba 
\{E{\scriptstyle (\a)}^j_A , J{\scriptstyle (\a)}_B \}&=& f_{AB}^{\ \
D}E{\scriptstyle (\a)}^j_D \label{}\\ 
\{M{\scriptstyle (\a)}_{jA}^{\ \ C} , J{\scriptstyle (\a)}_B \}&=&
f_{AB}^{\ \ D}M{\scriptstyle (\a)}_{jD}^{\ \ C} 
\ea 

\ni
and the flatness constraint generates ``translations'' of ${\bf E}$ 

\ba 
\{E{\scriptstyle (\a)}^j_A , P{\scriptstyle (\a)}_{jk}^B \} = \delta
_A^{\ B} + O(P) \quad . 
\ea 

\ni
where in $O(P)$ I group a collection of terms of first and higher
order in 
the curvature. Along the paper I am going to keep track of terms 
that vanish 
in this model where the lattice is flat, in order to be able to
discuss the 
issue of extending the model to a theory for general lattices. 

A remarkable feature of the Poisson algebra (\ref{EE})-(\ref{EM^-1}), 
and hence of the constraints (\ref{cerra}), (\ref{plani}), is that
under a 
decomposition of the variables into their self-dual and antiself-dual 
parts, the whole theory splits into two identical parts related by
complex conjugation. 

\subsection{Self-Dual Variables} 

 In $SO(3,1)$  apart from the Cartan metric $g_{AB}$, there is another 
invariant symmetric bilinear 
form $g^*_{AB}=g^*_{[ab][cd]}:=\varepsilon _{[ab][cd]}$. 
Its invariance follows directly from the invariance of the 
four-volume element under Lorentz transformations. 
This metric is used to define duality in the Lie algebra 

\be
V^*_A=g^{*B}_A V_B \quad ,\quad  g^{*B}_A:=g^*_{AC}g^{CB} \quad .
\ee

\ni 
The Lorentzian signature of the Cartan metric implies 
$\ g^{\ast B}_A g^{\ast C}_B =-\delta _A^{\ C}$. Therefore, to 
split them {\em real} Lorentz Lie algebra 
into its self-dual $^{(+)}$ and anti-self-dual $^{(-)}$ components, 
the projectors%
\footnote{These are projectors of the complexified Lie algebra. Here, one  
first includes the real Lie algebra into the 
complex Lie algebra 
and then split it into its self and antiself-dual parts. The fact that 
the images of $\d ^{(\pm)}$ lie out side of the image of the real Lie 
algebra does not prevent the ``split''; the only objection could be 
to call $\d ^{(\pm)}$ projectors.
} 
involve complex numbers. 

\ba
V^{(\pm)}_A:=\d ^{(\pm)B}_AV_B  
&=& \frac{1}{2}(\d ^B_A\mp i g^{\ast B}_A)V_B \\
g^{\ast B}_AV^{(\pm)}_B  
&=& \pm i V^{(\pm)}_A \quad .
\ea

The images of the self-dual and antiself-dual projectors are 
complementary orthogonal subspaces of $so(3,1;C)$. Also, 
the following formulas containing the structure constants hold 

\ba 
\d ^{(\pm)B}_A f_{BCD} = 
\d ^{(\pm)B}_A\d ^{(\pm)E}_C f_{BED}&=&
\d ^{(\pm)B}_A
\d ^{(\pm)E}_C\d ^{(\pm)F}_D  f_{BEF}:=f^{(\pm)}_{ACD} \\
\d ^{(+)B}_A\d ^{(-)E}_C f_{BED}&=&0 \\
f^{(+)}_{ABC}+f^{(-)}_{ABC} &=& f_{ABC} \quad .
\ea

\ni
In terms of self and antiself-dual variables 

\ba
E{\scriptstyle (\a)}^{(\pm)j}_A
&:=& \d ^{(\pm)C}_A E{\scriptstyle (\a)}^j_C \\
M{\scriptstyle (\a)}^{(\pm)B}_{jA}
&:=& \d ^{(+)C}_A \d ^{(\pm)B}_D M{\scriptstyle (\a)}^{\ \ D}_{jC}=
\d ^{(\pm)C}_A M{\scriptstyle (\a)}^{(\pm)B}_{jC} \quad ,
\ea

\ni
the Poisson algebra is 

\ba 
\{E{\scriptstyle (\a)}^{(\pm)\b}_A , E{\scriptstyle (\a)}^{(\pm)\b}_B \}
&=& f_{AB}^{\ \ D}E{\scriptstyle (\a)}^{(\pm)\b}_D 
= \pm f_{AB}^{(\pm)\ D}E{\scriptstyle (\a)}^{(\pm)\b}_D 
\label{EEpm}\\
\{E{\scriptstyle (\a)}^{(\pm)\b}_A , 
M{\scriptstyle (\a)}_{\b\ B}^{(\pm)C} \}
&=& f_{AB}^{\ \ D}M{\scriptstyle (\a)}_{\b \ D}^{(\pm)C} 
= \pm f_{AB}^{(\pm)\ D}M{\scriptstyle (\a)}_{\b \ D}^{(\pm)C} 
\label{EMpm}\\
\{E{\scriptstyle (\a)}^{(\pm)\b}_A , 
M{\scriptstyle (\b)}_{\a \ B}^{(\pm)C} \}
&=& f_{A\ D}^{\ C}M{\scriptstyle (\b)}_{\a \ B}^{(\pm)D} 
= \pm f_{A\ \ D}^{(\pm)C}M{\scriptstyle (\b)}_{\a \ B}^{(\pm)D} 
\label{EM^-1pm} \quad .
\ea

\ni 
The Poisson brackets between self-dual and antiself-dual variables always 
vanish. Since the structure constants $f^{(+)C}_{AB}$ and $f^{(-)C}_{AB}$ 
are totally antisymmetric three tensors in three-dimensional 
(complex) spaces, they are proportional to the intrinsic volume element. 
In the basis suggested by the reality conditions of next  section 
the proportionality constant for the self-dual part is $i\sqrt{2}$ 
and for the antiself-dual is $- i\sqrt{2}$. 
That is, the Lie algebra $so(3,1)$ ``splits'' into two 
copies of the Lie algebra $so(3;C)$. 
Each of these $so(3;C)$ algebras contains all the 
information of $so(3,1)$ 

\be 
V_A = V^{(+)}_A + V^{(-)}_A = V^{(+)}_A + c.c. \quad .
\ee 

\ni
An immediate but important consequence is that the symmetry generators 
also split, yielding two parallel theories. 

It would have been possible to start with self-dual variables in the 
action; however, 
I decided against it in order to preserve the direct geometric 
interpretation of the variables. On the other hand, using 
self-dual variables one learns that the geometricity conditions 
are the lattice counterpart of the reality conditions of Ashtekar's GR. 

\section{Geometricity-Reality Conditions}\label{geo-reali}

 The motivation for demanding geometricity conditions on the variables 
is to guarantee the existence of a one-to-one mapping between the 
space of simplicial lattices and the space of variables ${\bf E}$ , 
${\bf M}$ satisfying the geometricity conditions. Simultaneously, 
one gets a selection rule for the symmetry generators, ruling out 
the symmetries that do not preserve the geometricity conditions. 

A set of variables ${\bf E}{\scriptstyle (\a)}^j$ related to a face of 
a lattice of simplices is of the form 

\ba
E{\scriptstyle (\a)}^j_A =E{\scriptstyle (\a)}^j_{[ab]} 
&=& \frac{1}{2} \varepsilon_{abcd}\ 
l{\scriptstyle (\a ,j_1)}^c l{\scriptstyle (\a ,j_2)}^d \nl  
&=& \frac{1}{2} \varepsilon_{abcd}\ 
l{\scriptstyle (\a ,j_2)}^c l{\scriptstyle (\a ,j_3)}^d = 
\frac{1}{2} \varepsilon_{abcd}\ 
l{\scriptstyle (\a ,j_3)}^c l{\scriptstyle (\a ,j_1)}^d 
\label{e=ll}
\ea

\ni 
where the {\em space-like} Minkowski 
vectors ${\bf l}{\scriptstyle (\a ,j)}$ are associated with the 
links of the face that is the frontier 
between the cell $\a$ and its neighbor in direction $(j)$. 
Clearly, the link vectors satisfy the condition 
${\bf l}{\scriptstyle (\a ,j_1)}+{\bf l}{\scriptstyle (\a ,j_2)}+{\bf
l}{\scriptstyle (\a ,j_3)}=0$. 
Since every link of a tetrahedron is shared by two  of its 
faces, a relation of the 
form ${\bf l}{\scriptstyle (\a ,j_2)}=-{\bf l}{\scriptstyle (\a
,k_1)}$ holds for each link too. 

The geometricity conditions (\ref{e=ll}) are equivalent to the lattice 
analog of the condition $B = e \w e$ 
that distinguishes gravity from $B\wedge F$ theory. In this sense, 
discarding 
the symmetries that do not preserve the geometricity conditions bring
us an step closer to gravity. To avoid 
confusion between the $B$ of $B\w F$ theory and the magnetic field of 
the curvature in the lattice in the lattice I will write $b = e \w e$, 
more precisely, 

\ba 
E{\scriptstyle (\a)}^j_{[ab]} &\approx & \frac{1}{8} \hat{\ve}^{jkl}
b{\scriptstyle (\a)}_{kl[ab]}=  
\frac{1}{8} \hat{\ve}^{jkl} \ve _{abcd} e{\scriptstyle (\a)}^c_k
e{\scriptstyle (\a)}^d_l \nl  
&=&\frac{1}{32} \hat{\ve}^{jkl} \ve _{ab}^{\ \ cd} 
\hat{\ve}_{kmn} l{\scriptstyle (\a)}^{mn}_c \hat{\ve}_{lpq}
l{\scriptstyle (\a)}^{pq}_d 
\ea 
\ni 
where $l{\scriptstyle (\a)}^{jk\ c}= \hat{\ve}^{jkl} e{\scriptstyle
(\a)}^c_l$. The weak equivalence sign 
indicates that I have used the constraint 
$J{\scriptstyle (\a)}_A ={\sum _j} E{\scriptstyle (\a)}^j_A \approx
0$. I decided to write the 
``affine triads'' $e{\scriptstyle (\a)}^a_j$ that appear just as an
intermediate step between 
$E$'s and $l$'s to make contact with other works, and because these 
affine triads are the ones that indicate directions naturally in the 
lattice, and are going to be very helpful to write the constraints. 

All the geometricity requirements (\ref{e=ll}) for cell $(i)$ 
can be written purely in terms of the variables 
${\bf E}{\scriptstyle (\a)}^j$ 

\ba 
q{\scriptstyle (\a)}^{*jk} := g^{*AB}E{\scriptstyle
(\a)}^j_AE{\scriptstyle (\a)}^k_B = 0 \label{igeo} 
\ea 
\ni
or in terms of self-dual variables 
\be 
i(E{\scriptstyle (\a)}^{(+)jA}E{\scriptstyle (\a)}^{(+)k}_A - c.c. ) =
-2 \hbox{\rm Im}(E{\scriptstyle (\a)}^{(+)jA}
E{\scriptstyle (\a)}^{(+)k}_A) 
= 0 \quad .\label{qreal}
\ee

\ni 
This first set of conditions guarantees the geometricity 
of each separate cell: for $j=k=\b$ it requires that 
${\bf E}{\scriptstyle (\a)}^j$ represent the dual of the area bivector of 
the face  $(\a ,\b)$ between cells $\a$ and $\b$. 
In addition, for $j\neq k$, the condition is 
satisfied if the faces $(\a ,j)$ and $(\a ,k)$ of cell $\a$
intersect. Once conditions(\ref{qreal}) are satisfied, the variables 
${\bf E}{\scriptstyle (\a)}^j$ characterize a tetrahedron that is
contained in a 
{\em space-like} three-dimensional subspace of Minkowski space-time if 

\ba 
g^{AB}E{\scriptstyle (\a)}^j_AE{\scriptstyle (\a)}^j_B 
= 2 \hbox{\rm Re}(g^{AB}E{\scriptstyle (\a)}^{(+)j}_AE{\scriptstyle
(\a)}^{(+)j}_B) < 0 \label{<0} \quad . 
\ea 

The similarity between (\ref{qreal}) and the condition which requires the 
spatial metric of Ashtekar's formulation of gravity to be real 
is remarkable considering that the 
geometricity conditions (\ref{igeo}) were first proposed \cite{tlg} 
in a context not related to 
Ashtekar's formulation of general relativity. Furthermore, inequality 
(\ref{<0}) has a continuum analog that 
demands the metric to be Lorentzian. 

One also wants a covariant description in which parallel transport 
between neighboring faces is described by Lorentz matrices 
$M{\scriptstyle (\a)}_{j\ a}^{\ \ \ b}$ %
\footnote{I will use the same notation 
$M{\scriptstyle (\a)}_{j\ a}^{\ \ \ b} = \exp(A{\scriptstyle
(\a)}_j^{cb}\eta _{ca})$ 
for these matrices, which act on Minkowski vectors, 
as for the previously defined 
$M{\scriptstyle (\a)}_{jA}^{\ \ B} = \exp(A{\scriptstyle (\a)}_j^C
f_{CA}^{\ \ B})$ 
in the bivector representation. Both matrices are different 
representations of the same $SO(3,1)$ element. 
}. 
The variables ${\bf l}$ and ${\bf M}$ are called the geometrical 
variables. After enforcing the first set of geometricity conditions 
(\ref{qreal}), the geometrical variables are completely determined by the 
fundamental variables ${\bf E}, {\bf M}$. 
The complete set of geometricity 
conditions for the fundamental variables must imply that 
the geometrical variables of cell $\a$ and its neighbor $\b$ 
satisfy the compatibility conditions 

\be
{\bf l}{\scriptstyle (\a)}^{\b \c}=-{\bf M}{\scriptstyle
(\a)}_{\b}{\bf l}{\scriptstyle (\b)}^{\a \f} 
\label{l=-ml}
\ee 

\ni
where the faces defining ${\bf l}{\scriptstyle (\a)}^{\b \c}$ are
${\bf E}{\scriptstyle (\a)}^{\b}$, ${\bf E}{\scriptstyle
(\a)}^{\c}$,  and the ones defining 
${\bf l}{\scriptstyle (\b)}^{\a \f}$ are
${\bf E}{\scriptstyle (\b)}^{\a}$, ${\bf E}{\scriptstyle
(\b)}^{\f}$.  Some of these requirements are contained in the 
identity ${\bf E}{\scriptstyle (\a)}^{\b}=-{\bf M}{\scriptstyle
(\a)}_{\b} {\bf E}{\scriptstyle (\b)}^{\a}$; 
however, other conditions exist. These new restrictions 
relate to the different ways in which 
${\bf E}^*$ can be written as ${\bf l}\wedge {\bf l}$; 
in particular, there is one degree of freedom 
corresponding to the rotations within the plane defined by ${\bf E}^*$. 
The constraint on the connection matrices, which freezes this degree of 
freedom, imposes {\em zero torsion} on the lattice. 

An expression of condition (\ref{l=-ml}) in terms of the 
fundamental variables ${\bf E}, {\bf M}$ follows from its 
geometrical meaning. 
The condition requires that links on the boundary between 
cells $\a$ and $\b$ be the same when defined using variables 
${\bf E}$ from either cell. 
Consider ${\bf l}{\scriptstyle (\a)}^{\b \c}$ lying in 
the intersection of faces ${\bf E}{\scriptstyle (\a)}^{*{\b}}$ and 
${\bf E}{\scriptstyle (\a)}^{*\c}$ of cell 
$\a$ and ${\bf l}{\scriptstyle (\b)}^{\a {\f}}$ lying in the intersection 
of two faces ${\bf E}{\scriptstyle (\b)}^{*{\a}}$ and ${\bf
E}{\scriptstyle (\b)}^{*{\f}}$ of cell $\b$. 
Then equation (\ref{l=-ml}) holds 
if the planes defined by 
${\bf E}{\scriptstyle (\a)}^{*{\b}}$, ${\bf E}{\scriptstyle
(\a)}^{*\c}$ and ${\bf M}{\scriptstyle (\a)}_{\b} {\bf E}{\scriptstyle
(\b)}^{*{\f}}$ 
all intersect (\ref{g*ee}), (\ref{g*eme1}), (\ref{g*eme2}) and intersect 
along the same line (\ref{f*eeme}): 

\ba 
&&g^{*AB}E{\scriptstyle (\a)}^{\b}_AE{\scriptstyle (\a)}^\c_B = 
-2 \hbox{\rm Im}(E{\scriptstyle (\a)}^{(+){\b}B}E{\scriptstyle
(\a)}^{(+)\c}_B) =0 \label{g*ee} \\ 
&&g^{*AB}E{\scriptstyle (\a)}^{\b}_A M{\scriptstyle (\a)}_{{\b}B}^{\ \
C}E{\scriptstyle (\b)}^{\f}_C =  
-2 \hbox{\rm Im}(E{\scriptstyle (\a)}^{(+){\b}B} M{\scriptstyle
(\a)}_{{\b}B}^{(+)C}E{\scriptstyle (\b)}^{(+){\f}}_C)  
=0 \label{g*eme1} \\
&&g^{*AB}E{\scriptstyle (\a)}^\c_A M{\scriptstyle (\a)}_{{\b}B}^{\ \
C}E{\scriptstyle (\b)}^{\f}_C = 
-2 \hbox{\rm Im}(E{\scriptstyle (\a)}^{(+)\c B}M{\scriptstyle
(\a)}_{{\b}B}^{(+)C}E{\scriptstyle (\b)}^{(+){\f}}_C) 
=0 \label{g*eme2} \\
&&E{\scriptstyle (\a)}^{*\underline{{\b}}}_A f^{ABC}
E{\scriptstyle (\a)}^{*\c}_B M{\scriptstyle
(\a)}_{\underline{{\b}}C}^{\ \ D} E{\scriptstyle (\b)}^{*{\f}}_D \nl 
&&=2 \hbox{\rm Im}(E{\scriptstyle (\a)}^{(+)
\underline{{\b}}}_A f^{(+)ABC}
E{\scriptstyle (\a)}^{(+)\c}_B M{\scriptstyle
(\a)}_{\underline{{\b}}C}^{(+)D}E{\scriptstyle (\b)}^{(+){\f}}_D) 
=0 \quad  \label{f*eeme}
\ea

\ni
where there is no summation over the underlined index 
$\underline{{\b}}$ of the last equation. 
Condition (\ref{g*ee}) was already present in the first set of 
geometricity conditions on cell $\a$ (\ref{qreal}), and condition 
(\ref{g*eme1}) is a consequence of the geometricity 
conditions for cell 
$\b$ and the identity (\ref{anti}). The second set of geometricity 
conditions is: one condition of the kind (\ref{g*eme2}) and 
one condition of the kind (\ref{f*eeme}) for the relation 
${\bf l}{\scriptstyle (\a)}^{\b \c}=-{\bf M}{\scriptstyle
(\a)}_{\b}{\bf l}{\scriptstyle (\b)}^{\a \f}$ and other 
set of conditions for the relation 
${\bf l}{\scriptstyle (\a)}^{\b \c}=-{\bf M}{\scriptstyle
(\a)}_{\c}{\bf l}{\scriptstyle (\c)}^{\th \a}$. 
Clearly, the relations arising from 
${\bf l}{\scriptstyle (\a)}^{\b \c}=-{\bf M}{\scriptstyle
(\a)}_{\b}{\bf l}{\scriptstyle (\b)}^{\a \f}$ 
and the ones coming from 
${\bf l}{\scriptstyle (\b)}^{\a \f}=-{\bf M}{\scriptstyle
(\b)}_{\a}{\bf l}{\scriptstyle (\a)}^{\b \c}$ are equivalent; 
then an straight forward exercise in linear algebra shows that the 
``symmetrized'' relations 

\ba 
&&\hbox{\rm Im}(E{\scriptstyle (\a)}^{(+)\c B}M{\scriptstyle
(\a)}_{{\b}B}^{(+)C}E{\scriptstyle (\b)}^{(+){\f}}_C) + 
\hbox{\rm Im}(E{\scriptstyle (\a)}^{(+)\b B}M{\scriptstyle
(\a)}_{{\c}B}^{(+)C}E{\scriptstyle (\c)}^{(+){\th}}_C) 
=0 \label{sg*eme2} \\
&&\hbox{\rm Im}(E{\scriptstyle (\a)}^{(+)\underline{{\b}}}_A 
f^{(+)ABC}
E{\scriptstyle (\a)}^{(+)\c}_B M{\scriptstyle
(\a)}_{\underline{{\b}}C}^{(+)D}E{\scriptstyle (\b)}^{(+){\f}}_D) \nl
&&+ \hbox{\rm Im}(E{\scriptstyle (\a)}^{(+)
\underline{{\c}}}_A f^{(+)ABC}
E{\scriptstyle (\a)}^{(+)\b}_B M{\scriptstyle
(\a)}_{\underline{{\c}}C}^{(+)D}E{\scriptstyle (\c)}^{(+){\th}}_D) 
=0 \quad  \label{sf*eeme}
\ea

\ni
imposed for every link of each cell are equivalent to relations
(\ref{g*eme2}), (\ref{f*eeme}) imposed once for each of the 
three links of each face of every cell. 

In a recent paper, 
Immirzi wrote the reality conditions (\ref{qreal}) for the lattice; 
the motivation of his work is to obtain a 
consistent Ashtekar-like framework for the lattice \cite{immirzi}. 
This coincidence with Immirzi's formulas, that come from a rather 
different 
approach, is the lattice manifestation of a well known result 
of Capovilla, 
et al \cite{capovilla}. The mentioned result arrives to the Ashtekar 
formalism as the Hamiltonian version of a formulation of gravity 
based on two forms. 

The set of geometricity conditions (\ref{qreal}), (\ref{sg*eme2}), 
(\ref{sf*eeme}) on the fundamental variables form a {\em complete} set 
in the 
sense that they imply the existence and uniqueness of the geometrical 
variables (\ref{e=ll}) and that condition (\ref{l=-ml}) holds. 
An expression for the link $l{\scriptstyle (\a)}^{\c \d \ a}$ 
of tetrahedron $\a $ where face $(\a \c )$ 
($E{\scriptstyle (\a)}^{\c}_A =E{\scriptstyle (\a)}^{\c}_{[ab]} = 
\frac{1}{2} 
\varepsilon_{ab}{\ \ cd}\ l{\scriptstyle (\a)}^{\c \d }_c
l{\scriptstyle (\a)}^{\c \b }_d$), and 
face $(\a \d )$ 
($E{\scriptstyle (\a)}^{\d}_A =E{\scriptstyle (\a)}^{\d}_{[ab]} = 
\frac{1}{2} 
\ve_{ab}^{\ \ cd} l{\scriptstyle (\a)}^{\d \b }_c l^{\d \c }_d$) 
intersect is  

\be 
l{\scriptstyle (\a)}^{\c \d }_a=\frac{1}{v{\scriptstyle (\a)}}
\ve {\scriptstyle (\a)}_{abc} \ve {\scriptstyle (\a)}^{bef}
E{\scriptstyle (\a)}^{*\c }_{ef} 
\ve {\scriptstyle (\a)}^{cgh} E{\scriptstyle (\a)}^{*\d }_{gh} \quad ,
\label{l}
\ee
\ni
where the volume of the tetrahedron $\a $ is given by 
\be 
v{\scriptstyle (\a)}^2 = \frac{1}{9} 
\hat{\ve}_{jkl} f^{ABC} E{\scriptstyle (\a)}^{*j}_A E{\scriptstyle
(\a)}^{*k}_B E{\scriptstyle (\a)}^{*l}_C 
\ee 
\ni
and the volume element for cell $\a $ 
\ba
&&\ve {\scriptstyle (\a)}_{abc}={\phi {\scriptstyle (\a)}_{abc}
\over \sqrt{
{1\over 6} \phi {\scriptstyle (\a)}_{def} 
\phi {\scriptstyle (\a)}^{def}}} \\
&&\phi {\scriptstyle (\a)}_{abc}= \hat{\ve}_{jkl} \ve _{defg} 
E{\scriptstyle (\a)}_{\ \ [a}^{*j\ \ d} E{\scriptstyle (\a)}_{\ \
b}^{*k\ e} E{\scriptstyle (\a)}_{\ \ c]}^{*l\ \ f} 
x{\scriptstyle (\a)}^g 
\ea
\ni
does not depend on the choice of the transverse vector $x{\scriptstyle
(\a)}^g$. 
The geometricity conditions (\ref{qreal}), (\ref{g*eme2}), 
(\ref{f*eeme}) are also 
{\em necessary}: any set of variables ${\bf E}$, that can 
be written as in (\ref{e=ll}) where the compatibility condition 
(\ref{l=-ml}) holds, satisfies them.

>From (\ref{l=-ml}) one can see that if 
${\bf W}{\scriptstyle (\a)}_{\b \c }:=
({\bf M}{\scriptstyle (\a)}_{\b}M{\scriptstyle (\b)}_{\m} \ldots
M{\scriptstyle (\n)}_{\c} M{\scriptstyle (\c)}_{\a})$ is the 
holonomy around ${\bf l}{\scriptstyle (\a)}^{\b \c}$, 
then a consequence of the geometricity conditions is 
\ba 
{\bf l}{\scriptstyle (\a)}^{\b \c} = {\bf W}{\scriptstyle (\a)}_{\b
\c} {\bf l}{\scriptstyle (\a)}^{\b \c} 
\ea 
\ni
Because the geometricity conditions imply that ${\bf W}{\scriptstyle
(\a)}_{\b \c }$ has an 
axis and that the axis cross the planes of ${\bf E}{\scriptstyle
(\a)}^{\b *}$ and 
${\bf E}{\scriptstyle (\a)}^{\c *}$
\ba 
P{\scriptstyle (\a)}_{\b \c}^{*A} P{\scriptstyle (\a)}_{\b \c A} 
&=& -2\hbox{\rm Im}(P{\scriptstyle (\a)}_{\b \c}^{(+)*A} P{\scriptstyle
(\a)}_{\b \c A}^{(+)}) = 
0 \label{p*p} \\
P{\scriptstyle (\a)}_{\b \c}^{*A} E{\scriptstyle (\a)}^{\b}_A  
&=& -2\hbox{\rm Im}(P{\scriptstyle (\a)}_{\b \c}^{(+)*A} E{\scriptstyle
(\a)}^{(+)\b}_A) = 0 \label{p*e} \\
E{\scriptstyle (\a)}^{\b *}_A f^{AB}_{\ \ C}E{\scriptstyle
(\a)}^{\c}_B P{\scriptstyle (\a)}_{\b \c}^{C}   
&=& 2\hbox{\rm Im}(E{\scriptstyle (\a)}^{(+)\b *}_A f^{AB}_{\ \ C}
E{\scriptstyle(\a)}^{(+)\c}_B  
P{\scriptstyle (\a)}_{\b \c}^{(+)C}) = 
0 \label{e*fep} \quad .
\ea 

Evidently the geometricity conditions in the lattice are stronger than 
what one expected from experience in the continuum; 
in particular, the continuum counterpart of 
relations (\ref{p*e}), (\ref{e*fep}) does not hold. This 
rigidity of the lattice makes some of the constraints trivial. 
I discuss this issue in the next two sections.  

\section{The Constraints}\label{constraints}

 Not all the symmetries generated by the constraints 
$P\hskip -.05cm \approx \hskip -.05cm 0$ 
of lattice $B\hskip -.03cm {\w}\hskip -.03cm F$ theory 
(LBF) preserve the geometricity conditions. 
The largest subgroup that preserves them is the group of translation 
of vertices of the lattice. The generator of translations of vertex 
$(v)$ of cell $\a $, in the direction of $w(\a)^{(+)a}$, was 
introduced in \cite{tlg}. It is easily written in self-dual variables 
using that $\d ^{(+)}= {1\over 2}(g-ig^*)$ and that $w(\a)^{(+)a}$ 
is real
 
\ba 
2 \hbox{\rm Re}(w(\a)^{(+)a} T(\a ,v)^{(+)}_a) &:=& 
\bigg[ w(\a)^{(+)a} \sum_{\{ jk\} \to v} 
\frac{1}{2} \ve _{abcd} l(\a)^{jk\ b} 
\al P_{jk}^{(+)cd} \\ 
&+& (M(\b)_{\a} w(\a)^{(+)})^a 
\sum_{\{ jk\} ^-\to v} \frac{1}{2} \ve _{abcd} 
l(\b)^{jk\ b} P(\b)_{jk}^{(+)cd} + \ldots \bigg] + c.c. 
\nonumber \label{Tv}
\ea 

\ni
The summation written above runs over the links $l(\a)^{jk\ a}$ 
pointing in the direction of vertex $(v)$. 
The terms of the summation have been split for convenience, according to 
the cell $\a ,\b , \ldots $ where the variables are expressed, but each 
link must be included only once in the summation. For this reason, 
the index of the second summation written is $\{ jk\} ^-$ indicating to 
sum only over the links not included in the previous summation. 
 
One can easily prove that the action of 
$T(\a ,v)_a:=2\hbox{\rm Re}(T(\a ,v)^{(+)}_a)$ 
on the variables $E$, that determine the geometricity of the lattice, 
is to generate translations of the vertex $(v)$. First one sees that if 
the face $\s$ of cell $\r$ does not contain vertex $(v)$ 
the Poisson bracket of 
$T(\a ,v)_a$ and $E(\r)^{\s}$ is zero; and that 
in the case of a face that contains $(v)$, 
like face $\b$ of cell $\a$, the action of the generator 
$T(\a ,v)_a$ on the variable $E(\a)^{*\b}_ A = E(\a)^{*\b}_{[ab]} 
= l(\a)^{\b \c }_{[a} l(\a)^{\b \d }_{b]} = 
l(\a)^{\b \d }_{[a} l(\a)^{\b \e }_{b]} = 
l(\a)^{\b \e }_{[a} l(\a)^{\b \c }_{b]}$ 
has the same geometrical effect as a translation of the vertex 
$(v)$, where faces $\b$, $\c$, $\d$ of cell $\a$ intersect.  

\ba 
\{ E(\a)^{*\b}_{ef} , w(\a)^{(+)}_a T(\a ,v)^a \} 
&= &\frac{1}{4} \varepsilon ^{ab}_{cd} \varepsilon _{ef}^{gh}
w(\a)^{(+)}_a l(\a)^{\b \c}_b \{ E(\a)^{\b }_{gh} , 
P(\a)_{\b \c}^{cd} \} \nl 
&+&\frac{1}{4} \varepsilon ^{ab}_{cd} \varepsilon _{ef}^{gh}
w(\a)^{(+)}_a l(\a)^{\b \d}_b \{ E(\a)^{\b }_{gh} , 
P(\a)_{\b \d}^{cd} \} + O(P) \nl
&=& w(\a)^{(+)}_{[e} ( l(\a)^{\b \c}_{f]}  - 
l(\a)^{\b \d}_b ) + O(P) \nl 
&=& 
w(\a)^{(+)}_{[e}  l(\a)^{\b \e}_{f]} 
+ O(P) \quad . \label{Tr}
\ea 

\ni
An immediate consequence is that the geometricity conditions 
are preserved by the translation generators 
\be 
\{ q(\b)^{*jk} , w(\a)^{(+)a} T(\a ,v)_a \} = 0 + O(P) 
\quad . \label{repres} 
\ee 

Two remarks are in order. First, 
even though it seems natural to consider 
$T(\a ,v)_a$ as the symmetry generator but leave the full (complex) 
$T(\a ,v)^{(+)}_a \approx 0$ as constraint, with the reality condition 
$w(\a)^{(+)a}\epsilon R$ and a real Hamiltonian, 
the model would not make sense 
without the geometricity conditions. 
The strategy of forging a consistent 
theory for lattice gravity even without geometricity conditions 
is discussed in the last section. 
Second,  $\hbox{\rm Im}(T(\a ,v)^{(+)}_a) = 0$ is automatically 
satisfied for lattices where  the geometricity conditions are 
satisfied; thus, writing the constraints 
as $T(\a ,v)^{(+)}_a \approx 0$ {\em is correct}, but only half of them 
are independent of the geometricity conditions (and the formula makes 
sense only for geometrical lattices). 
 
The result of imposing the symmetries generated by 
$T(\a ,v)_a$ is a theory (GLBF) that describes the 
geometric sector of LBF \cite{tlg}. More precisely: 

\begin{itemize} 
\item All the solutions of GLBF are solutions of LBF. 
\item The solutions of GLBF are the flat space-times $\Sigma \times R$ 
generated by a geometrical lattice (a lattice made of vertices, links, 
faces and cells) $\Sigma $ during its evolution. 
\item There are some solutions of LBF with ``global torsion'' that 
do not admit any geometric representation \cite{waelbroeckBF}. 
\item Both 
GLBF and LBF have {\em zero} local degrees of freedom. Both theories 
have only discrete, topological, degrees of freedom. 
\item GBF, the restriction of $B\w F$ to torsion free connections is 
equivalent to GLBF. The proof follows the procedure used by Waelbroeck to 
prove the equivalence of the lattice $2+1$ theory and the continuum 
theory 
\cite{waelbroeckPRL}
\end{itemize}

A lattice theory for gravity must be geometrical (at least when 
restricted to flat space-times), and must possess local degrees of 
freedom to reproduce a theory with two degrees of freedom per 
point in its macroscopic limit. 
The facts that GLBF has the largest symmetry group 
that preserves the geometricity of the lattice and 
has zero local degrees of freedom means that GLBF has too many 
symmetries. The next step in obtaining a lattice theory for 
gravity is therefore to select the correct subgroup of the symmetry 
group of  GLBF. The result of choosing the correct subgroup of the 
symmetry group of GLBF is the model presented in this article 
(a precise definition of what I mean by {\em the model} 
will be given shortly). As the symmetry group is smaller 
than that of GLBF one can interpret the model as the result of 
restricting a 
theory of lattice gravity to act on flat initial data.  
This hypothetical theory of lattice gravity would be a theory with 
local degrees of freedom, and part of the motivation of 
this article is to learn as 
much as possible about the hypothetical theory from its 
restriction to flat 
initial data (for an extended discussion see the concluding section). 

A suggestion of which subgroup to consider comes from the affine 
notation employed in this article. 
For the affine notation, cells 
are the lattice counterpart of points in the continuum. 
I will show 
that the translations of lattice cells constitute a proper subgroup of 
GLBF's symmetry group (the group of vertex translations). 
The generator of translations of cell $\a $ in the direction of 
$w(\a)^{(+)a}$ 
is simply the one that moves the four vertices of $\a $ 
by $w(\a)^{(+)a}$. 

\ba 
w(\a)^{(+)a} T(\a)_a 
&:=&w(\a)^{(+)a} T(\a ,v_1)_a + w(\a)^{(+)a} T(\a ,v_2)_a \nl 
&+& 
w(\a)^{(+)a} T(\a ,v_3)_a + w(\a)^{(+)a} T(\a ,v_4)_a \label{T}
\ea 

The commutativity of cell translations inside the group of 
vertex translations 
is a simple consequence of the commutativity of vector addition. 
Hence the 
symmetry group of this model has dimension $4N_3$ (four times the 
number of 
lattice cells). In a lattice that admits local deformations 
that make it flat the number of vertices is 
(locally) bigger than the number of cells 
(see the appendix). 
As a result, the symmetry group of this model is a proper subgroup of 
that of GLBF. 

In the lattice the links have been described using internal Minkowski 
vector spaces that are related by parallel transport matrices; the 
same internal 
space has been used to point the directions of translation for the 
translation 
generators. These internal Minkowski spaces are just side products of 
the geometricity conditions and are not 
the internal spaces where the dynamical variables live. 
Because of this, 
it would be desirable to label the translation generators 
of the lattice using notions that are more compatible with the continuum. 
A general translation of a cell $\a $ can be specified by a {\em real} 
``lapse''  $N(\a)^{(+)}$ and a {\em real}
``shift'' $N(\a)^{(+)j}$ $j = \b , \c , \d ,\e $ (using the affine 
notation described in sec.~\ref{affine} 
$2\hbox{\rm Re}(N(\a)^{(+)j} {\cal H}(\a)^{(+)}_j + N(\a)^{(+)} {\cal
H}(\a)^{(+)})$.  The generator 
${\cal H}(\a)_{\b}:=2\hbox{\rm Re}({\cal H}(\a)^{(+)}_{j=\b})$
moves cell $\a $ in the direction of its neighbor $\b $, 
and ${\cal H}(\a):=2\hbox{\rm Re}({\cal H}(\a)^{(+)})$ translates cell 
$\a $ in the direction orthogonal to the 
three dimensional space that contains it 

\ba 
&&{\cal H}(\a)^{(+)}_{\b} := {1\over 16} (l(\a)^{\e \c } + 
l(\a)^{\c \d } + l(\a)^{\d \e })^a T(\a)^{(+)}_a \approx 0 
\label{ha} \\
&&{\cal H}(\a)^{(+)} := {1\over 2} l(\a)^{\e \c }_a  
l(\a)^{\c \d }_b  l(\a)^{\d \e }_c 
\ve ^{abcd} T^{(+)}_d \approx 0 \label{h} \\
&&J(\a)^{(+)}_A ={\sum _j}E(\a)^{(+)j}_A \approx 0 \label{j} \\
&&P(\a)_{jk}^{(+)A}(\hbox{\small initial}) = 0 \label{p} 
\ea 

\ni
I wrote the lattice Gauss law, and the condition restricting to 
flat lattices as initial data in order to precise what I mean by 
{\em the model}. The model's phase space is that of LBF described 
by variables $E^{(+)}, M^{(+)}$ and Poisson bracket structure 
(\ref{EEpm}), (\ref{EMpm}), (\ref{EM^-1pm}). 
It is subject to the geometricity conditions (\ref{qreal}), 
(\ref{sg*eme2}), (\ref{sf*eeme}), and to constraints 
(translation and gauge generators) (\ref{ha}), 
(\ref{h}), (\ref{j}). Also the model's initial data is restricted to flat 
lattices (\ref{p}). The initial data condition can also be written as 
$T(\a)^{(+)}_a(\hbox{\small initial}) = 0$, in a geometrical lattice both 
relations are equivalent. This is what makes GLBF a genuine theory for the 
geometric sector of LBF. For the model presented in this article 
${\cal H}(\a)^{(+)}_{\b}=0$, ${\cal H}(\a)^{(+)}=0$ do not imply 
$P(\a)_{jk}^{(+)A} = 0$ even for a geometrical lattice; this opens the 
possibility of considering the model as the restriction of a 
hypothetical theory for 
non flat lattices to the case of initial data satisfying 
$P(\a)_{jk}^{(+)A}(\hbox{\small initial}) = 0$. The translation generators 
${\cal H}(\a)^{h(+)}_{\b}=0$, ${\cal H}(\a)^{h(+)}=0$ of the 
of the hypothetical lattice theory when restricted to flat lattices have 
the form (\ref{ha}), (\ref{h}), i.e. 
${\cal H}(\a)^{h(+)}_{\b}={\cal H}(\a)^{(+)}_{\b}+O(P^2)$, 
${\cal H}(\a)^{h(+)}= {\cal H}(\a)^{(+)}+O(P^2)$. The fact that the 
symmetry group of the model is smaller than that of GLBF means that the 
hypothetical lattice theory that reduces to the model for flat initial
data is a theory with local degrees of freedom. 

As in the case of GLBF's constraints, 
one should regard only 
${\cal H}(\a)_{j}$, ${\cal H}(\a)$ as symmetry generators and 
consider all the ${\cal H}(\a)^{(+)}_{j}\approx 0$, 
${\cal H}(\a)^{(+)}\approx 0$ as constraints. 
However, they are not independent within themselves%
\footnote{
In a three dimensional simplicial lattice there are $6(N_1-N_0)=6N_3$ 
independent curvature variables $P(\a)_{jk}^A$ because of the Bianchi 
identities; thus, 
the $8N_3$ projections of them (${\cal H}(\a)^{(+)}_j$, 
${\cal H}(\a)^{(+)}$) can not be independent. 
}, 
and their imaginary part is a direct consequence of the geometricity 
conditions. 

Although it is difficult to calculate the algebra of the symmetry 
generators, 
there is a simple way to prove that their Poisson bracket vanishes 
weakly. The translation generators 
${\cal H}(\a)_j$ and ${\cal H}(\a)$ are scalars and, therefore, their 
Poisson brackets with the gauge transformation generator $\al J^{(+)}_A$ 
vanish. These translation generators, (${\cal H}(\a)_j$ and 
${\cal H}(\a)$), 
span the same space as the former translation generators 
$T(\a)_a$. Introducing a bit of notation, 
$N(\a)^{(+)\m} {\cal H}(\a)_{\m}:=
N(\a)^{(+)j} {\cal H}(\a)_j + N(\a)^{(+)}{\cal H}(\a)$,  
the previous statement signifies that 
$T(\a)_a = C(a)_a^{\ \m} {\cal H}(\a)_{\m}$ 
where the matrix 
$C(a)_a^{\ \m}$ has rank four. Then the Poisson brackets of $T$'s and 
those of $H$'s are related by 

\ba 
\{ T(\a)_a , T(\b)_b \} &=& C(\a)_a^{\ \m}  C(\b)_b^{\ \n}  
\{ {\cal H}(\a)_{\m} , {\cal H}(\b)_{\n} \} \nl 
&+& C(\a)_a^{\ \m} \{ {\cal H}(\a)_{\m} , C(\b)_b^{\ \n}   \} 
{\cal H}(\b)_{\n} + {\cal H}(\a)_{\m} \{ C(\a)_a^{\ \m} , 
{\cal H}(\b)_{\n} \} 
C(\b)_b^{\ \n} \nl 
&\approx & C(\a)_a^{\ \m}  C(\b)_b^{\ \n} 
\{ {\cal H}(\a)_{\m} , {\cal H}(\b)_{\n} \} \quad . 
\ea 

\ni
An immediate consequence of the fact that 
the translation generators $T(\a)^{(+)}_a$ commute (in flat space-time) 
$\{ T(\a)_a , T(\b)_b \} = 0+O(P)$ is that 
the Poisson brackets of the new form of the translation generators 
weakly vanish up to second order in the curvature

\ba 
\{ {\cal H}(\a)_{\m} , {\cal H}(\b)_{\n} \} \approx 
0 + O(P^2) \quad . \label{[h,h]} 
\ea 

An extremely interesting feature of the spatial translation constraints 
${\cal H}(\a)^{(+)}_j$ and the time translation constraints 
${\cal H}(\a)^{(+)}$ 
is that their local parts are algebraically identical to the 
diffeomorphism and Hamiltonian constraints of 
the Ashtekar formulation of general relativity. To define the local parts 
one requires 

\ba 
&&\{ E(\a)^{(+)j}_A , {\cal H}(\a)^{(+)}_{\m} \} = 
\{ E(\a)^{(+)j}_A , {\cal H}(\a)^{(+)local}_{\m} \} \\
&&\{ M(\a)_{jA}^{(+)B} , {\cal H}(\a)^{(+)}_{\m} \} = 
\{ M(\a)_{jA}^{(+)B} , {\cal H}(\a)^{(+)local}_{\m} \} \quad .
\ea 
\ni
and that in the expression for ${\cal H}(\a)^{(+)local}_{\m}$ only 
variables related to links of cell $\a $ appear. The link variables 
$l(\a)^{jk\ a}$ occur in pairs that recombine in the form of the 
variables of the theory $E(\a)^{(+)j}_A$ to yield 

\ba 
{\cal H}(\a)^{(+)local}_j \hskip -.2cm 
&=&\hskip -.2cm {1\over 2}\hat{\ve}_{jkl} E(\a)^{(+)k}_A B(\a)^{(+)lA} 
- ({1\over 2} P(\a)^{(+)A}_{jk} \hat{n}^k)  J(\a)^{(+)}_A
\label{hal}\\
{\cal H}(\a)^{(+)local} \hskip -.2cm 
&=&\hskip -.2cm {1\over 4} \hat{\ve}_{jkl} 
f^{AB}_{\ \ C} E(\a)^{(+)j}_A E(\a)^{(+)k}_B 
B(\a)^{(+)C} \quad , \label{hl}
\ea 

\ni
where the ``magnetic field'' is written as 
$B(\a)^{(+)jA}:=\hat{\ve}^{jkl}P(\a)_{kl}^{(+)A}$. The second term of 
${\cal H}(\a)^{(+)local}_j$ generates Lorentz transformations and 
vanishes for flat space-times. 
The non-locality of the translation constraints 
${\cal H}(\a)^{(+)}_{\m}$ is very mild. 
The difference between ${\cal H}(\a)^{(+)local}_{\m}$ and 
${\cal H}(\a)^{(+)}_{\m}$ 
is function only of variables of the lattice sharing vertices with
cell $\a $. 
One of the first basic distinctions between the lattice and the continuum%
\footnote{
I am assuming that the continuum is an smooth manifold; 
obviously, excluding pathological topologies such as non-Hausdorff spaces. 
} 
appears: the concept of neighborhood fundamentally differs. 
Two points in the continuum $p$ and $q$ can either be the same or 
be separated by open neighborhoods. In contrast, two cells in the lattice 
can either be the same, be immediate neighbors, or be separated by 
other cells. Clearly the category of immediate 
neighbors disappears during any acceptable continuum limit; therefore,
in any 
acceptable (see last section for a discussion) continuum limit, only the 
local parts of the expressions are going to remain. Hence, the continuum 
limit of the translation generators is, precisely, Ashtekar's 
diffeomorphism and Hamiltonian constraints. 

In order to summarize ideas and prepare the discussion, I 
am going to count the degrees of freedom of the hypothetical lattice theory 
resulting from an extension of the constraints of the model (\ref{ha}), 
(\ref{h}) to first-class constraints ${\cal H}^h(\a)_j$, 
${\cal H}^h(\a)$. 
Recall that the number of points, links, faces, and cells are 
denoted by $N_0 ,N_1 ,N_2$ and $N_3$ respectively, and also, for 
a lattice of tetrahedra, $N_2 = 2 N_3$. 
The phase space variables are given by the $12 N_2 = 24 N_3$ numbers 
$E(\a)^{(+)j}_A$, $M(\a)_{jA}^{(+)B}$, that are subject to $6N_3$ closure 
constraints $J(\a)^{(+)}_A \approx 0$, and $3N_3$ vector constraints 
${\cal H}^h(\a)_j$, and $N_3$ scalar constraints 
${\cal H}^h(\a)$. These 
constraints are first-class and generate Lorentz transformations, 
spatial translations, and time translations respectively. Thus, 
the dimension of the reduced phase space, without taking into account the 
geometricity conditions, is 

\ba 
24 N_3 - 2 ( 6 N_3 + 3 N_3 + 1 N_3 ) = 4 N_3 \quad .
\ea 

\ni
And the geometricity conditions reduce the number of 
degrees of freedom to at least half. 
The reason is that all the scalar information in the variables 
$E(\a)^{(+)j}_A$of cell $\a $ 
is captured in the symmetric matrix
$q(\a)^{(+)jk}= E(\a)^{(+)j}_A E(\a)^{(+)k}_B g^{AB}$; 
thus, the first set of geometricty conditions 
($\hbox{\rm Im}(q(\a)^{(+)jk})=0$) 
reduces the scalar information on $E(\a)^{(+)j}_A$ from 
$12$ to $6$ numbers. Some local degrees of freedom remain, 
because as proven in the appendix, the 
symmetry group of the theory is ``locally smaller'' than that of GLBF, 
a theory with zero local degrees of freedom. For a discrete, 
{\em microscopic}, theory of gravity local degrees of freedom are 
essential, but to get the expected $2 N_3$ is not. 
A discussion of this point 
and related issues is the topic of the last section. 

\section{Discussion}

A classical 
lattice theory that describes space-time in $3+1$ form must be 
geometrical, 
i.e. a unique three-dimensional piecewise linear space must be
assigned 
to any 
set of variables of the theory (in the present case 
$\{ E, M\} $ satisfying the geometricity conditions, 
$\{ l, \pi _l \} $ in the usual formulation of Regge Calculus). 
Also, a lattice theory with no local degrees of freedom cannot be 
related to 
gravity simply because any sensible continuum limit cannot convert a set 
of configurations that are equivalent in the lattice description into 
inequivalent geometries in the corresponding continuum theory. 
On one hand, the model presented in this article is geometrical; 
on the other hand, however, the model describes flat space-times. In this 
sense the model is not better than GLBF (geometric lattice $B\w F$ theory). 
There are two essential differences between the model and GLBF. First,
GLBF is 
a theory with first-class constraints that describes the geometric sector 
of $B\w F$ theory in the sense described in section~\ref{constraints}. 
This is different from the model introduced in this article, that has a 
symmetry group whose flows commute only if the initial data is a flat 
lattice. 
Second, the constraints of the model do not force the lattice curvature to 
be zero; therefore one can consider the model as the restriction of a 
{\em hypothetical theory}, that includes non-flat lattices, restricted
to the 
case of flat initial data. 
The reason to call the theory ``hypothetical'' is not that its existence 
is in doubt; for instance, 
the method to find symplectic coordinates \cite{arnold} 
can be used to get an extension of 
the translation generators $T(\a ,v)_a$, in an open neighborhood of the 
submanifold $P=0$ where their flows commute, to momentum coordinate 
functions 
that with some configuration coordinate functions form a set of symplectic 
coordinates. Then, the extensions $T^h(\a ,v)_a$ of the translation 
generators of the model are the constraints of the hypothetical theory. 
The enormous freedom in the choice of coordinates makes 
this method more of an existence statement than a constructive process: 
that is the reason to call the extension h-theory.  
Along the article I presented four results concerning {\em any } 
h-lattice theory that reduces to the model when restricted to flat initial 
data. 

\begin{enumerate} 
\begin{enumerate} 
\item {\em There are local degrees of freedom in the h- theory}. This 
follows from the fact that the symmetry group of the model is a proper 
subgroup of that of GLBF (see appendix). 
\item {\em The continuum limit has to be a macroscopic limit}. 
If the continuum limit were taken by simply shrinking 
the lengths of the lattice links to zero, and by identifying the cells (or 
vertices) of the lattice with points of the continuum, the resulting 
continuum 
theory would have less degrees of freedom than gravity because the 
h-theory has less than two degrees of freedom per lattice cell. 
Thus, to be consistent one must regard this theory as a microscopic 
theory. 
\item {\em Regarding the macroscopic limit of the h-theory and Ashtekar's 
formulation of general relativity}. 
Although a serious study of the macroscopic limit is yet to be performed, 
two facts indicate that the macroscopic limit of the h-theory and general 
relativity formulated in terms of the new variables are related. 
First, the local part of the translation generators of the model is 
algebraically identical to Ashtekar's constraints, and the reality 
conditions 
have the geometricity-reality conditions as lattice counterpart. 
Second, an appropriate procedure to take the macroscopic limit 
could be through refinements of the (dual) 
lattice and a prescription for projecting down the the variables from the 
refined lattice to the original lattice. Along these refinements the 
phase space of the lattice becomes bigger and given a non-vanishing 
macroscopic curvature it is possible to reach it as the limit of lattices 
where the curvature in every link goes to zero. These special
``smooth  configurations'' would be the ones that define the spatial 
manifold 
$\Sigma$ and for them the constraints of the h-theory become the 
translation 
generators of the model. Furthermore, in the macroscopic limit the 
concept of 
immediate neighboring cells is lost and, hence, only the local part of the 
translation generators is relevant. This local part is the one that is 
algebraically identical to Ashtekar's constraints. 

\item {\em The h-theory is condemned to remain classical}. Immirzi's
observation \cite{immirziQUAN} that the quantum reality conditions a
nd the
algebra of the area bivectors is incompatible makes the theory well 
defined 
only classically (see discussion below). 

\end{enumerate} 
\end{enumerate} 

Most of the effort behind this article was directed towards a slightly
different goal than the one achieved. The relation between the
geometricity conditions and the reality conditions (when using
self-dual variables) was the motivation to structure a model for
lattice gravity that precisely mirrored Ashtekar's formulation of
GR. In particular, a self-consistent model with symmetry generators
whose flows commute for flat initial data ($P({\small initial})=0$),
that resembled Ashtekar's seemed feasible.  A summary of the results,
using as proposed constraints 
${\cal H}{\scriptstyle(\a)}^{(+)local}_{\m}$ 
(\ref{hal},\ref{hl}), is the following: 

\begin{itemize}
\item $\{ {\cal H}{\scriptstyle (\a)}^{local}_{\m} , 
{\cal H}{\scriptstyle (\b)}^{local}_{\n} \} \approx 0 + O(P^2)$ 
{\em except} in the case of $\a $ and $\b $ being neighbors sharing 
only a link. In this case $\{ {\cal H}{\scriptstyle (\a)}^{local}_{\m} , 
{\cal H}{\scriptstyle (\b)}^{local}_{\n} \} \approx 0 + O(P)$, which 
means that the 
symmetries generated by the constraints do not commute even for flat 
space-times. 
\item The ``symmetry generators'' 
${\cal H}{\scriptstyle (\a)}^{local}_{\m}$ do not respect 
the geometricity of neighboring cells even for flat space-times. 
\item On the other hand, the generators 
\ba 
{\cal H}{\scriptstyle (\a)}^{local}_j  
&=& 2\hbox{\rm Re}({1\over 2}\hat{\ve}_{jkl} E{\scriptstyle (\a)}^{(+)k}_A
B{\scriptstyle (\a)}^{(+)lA} 
- ({1\over 2} P{\scriptstyle (\a)}^{(+)A}_{jk} \hat{n}^k) J{\scriptstyle
(\a)}^{(+)}_A)
=\hat{\d }_j^{\ k}P{\scriptstyle (\a)}_{kl}^A E{\scriptstyle (\a)}^l_A \\
{\cal H}{\scriptstyle (\a)}^{local}  
&=& 2\hbox{\rm Re}({1\over 4} \hat{\ve}_{jkl} 
f^{AB}_{\ \ C} E{\scriptstyle (\a)}^{(+)j}_A E{\scriptstyle (\a)}^{(+)k}_B 
B{\scriptstyle (\a)}^{(+)C})=f^{AB}_{\ \ C}E{\scriptstyle (\a)}^j_A
E{\scriptstyle
(\a)}^k_B P{\scriptstyle (\a)}_{jk}^C
\ea 
\ni
{\em do} generate the translations expected for variables 
$E{\scriptstyle (\a)}^{(+)j}_A$ {\em of the same cell}. 
\end{itemize} 

\ni
The conclusion is that the constraints ${\cal H}{\scriptstyle
(\a)}^{(+)local}_{\m}$ are not correct even at first-order in the 
curvature; but they are the local part of 
constraints that are correct at first order. 
This attempt  deviates from the path that uses the lattice $B\w F$ 
theory as a firm 
ground from which one can extract a richer theory. Using only the local 
parts of the expressions written in this paper leads to a discrete 
theory closely related to a discretization of general relativity 
using Ashtekar's new variables proposed recently by 
Bostr\"{o}m, Miller and Smolin 
\cite{new-discre}. The translation generators 
introduced in this paper should be considered as a refinement of those 
proposed in \cite{new-discre}, because they differ only in their non-local 
part that is necessary for them to have commuting flows (for initial data 
satisfying $P=0$). In this sense, it is the first time that a connection 
oriented formulation of $3+1$ lattice gravity has symmetry generators that 
are correct up to first order. Thus, this model is of potential
interest to 
develop numerical codes. As shown in 
section~\ref{constraints} these non-local terms have an entirely
discrete 
origin and, therefore, their absentia is natural 
in a theory derived directly 
from a discrete analog of a continuum action. 
Unfortunately, only the local part is easily written 
in terms of the dynamical 
variables $E$, $M$; the non-local terms are naturally written using the 
geometrical variables $l$, $M$. The problem with straightforward
substitution using  the formulas giving the links 
$l{\scriptstyle (\a)}_{jk}^a $ in terms of the 
variables $E{\scriptstyle (\a)}^{(+)j}_A$ 
(\ref{l}) is that the 3-volume element 
$\ve {\scriptstyle (\a)} _{abc}$ for each tetrahedron 
$\a $ is a complicated function of the 
$E$'s. A solution to this problem could be modifying the 
model to have  $SO(3)$ as 
internal group and then inherit the natural volume element of the
Lie algebra. The counterpart of this approach in the continuum has 
no complicated reality conditions and is linked 
to Lorentzian gravity via the generalized Wick transform 
\cite{thiemannCOMPL,ashtekarWICK} while keeping the Hamiltonian
constraint algebraically simple. To import this approach to the
lattice means trading immediate geometrical interpretation for
algebraic simplicity. Direct geometric interpretation of the lattice
theory should not be regarded as a first priority, because a theory
of classical gravity with the correct number of degrees of freedom 
can be recovered only as the macroscopic limit of the lattice theory
anyway. 

An extension of the model that is coherent even without geometricity 
conditions could be the first step in extending the model to a
theory of lattice gravity that does not require 
$P(\hbox{initial})=0$. 
To pursue this option is particularly interesting, because 
the whole framework of lattices and projective techniques 
\cite{ALMMT,baez} is tailored for quantization. 

In an attempt to organize ideas for future investigations, a 
program towards a quantum  theory based on this kind of lattice 
gravity has been outlined. The consequences up to now 
have been only to state several well known facts in the language 
used in this article. As mentioned earlier, the lattice theory must
be regarded as a microscopic theory to achieve a continuum limit
with the correct number of degrees of freedom.  
A particularly appealing strategy is to do both, the connection to
continuum gravity and quantization, simultaneously; such a task 
is not an utopia, the projective techniques developed 
by Ashtekar et al \cite{ALMMT} and Baez 
\cite{baez} were designed for this purpose. An adaptation of the 
mentioned strategy for quantization has already been used for abstract 
lattices by Loll \cite{loll}. Immirzi showed that implementation of 
the quantum reality conditions that result from using $SL(2,C)$ or 
$SO(3,1)$ as internal groups is inconsistent; rather than a final 
word this observation 
should be considered as an other factor in 
favor of adopting Thiemann's strategy of solving the Lorentzian 
reality conditions via the generalized Wick transform. 
Work in this direction is in progress \cite{aa-ac-jz}. 

In the quantization program the issue of 
constructing a regularization of the constraints is a central one; 
provide hints to avoid future problems, 
like the presence of anomalies, was one of the motivations for 
working on this model. 

This work greatly benefited from 
enlightening discussions with Abhay Ashtekar, Alejandro Corichi, 
Jorge Pullin, Lee Smolin and Henri Waelbroeck. 
I would also like to thank Mary-Ann Hall for her editorial help. 
Support was provided by Universidad Nacional 
Aut\'onoma de M\'exico (DGAPA), and grants
NSF-PHY-9423950, NSF-PHY-9396246, research funds of 
the Pennsylvania State
University, the Eberly Family research fund at 
PSU and the Alfred P. Sloan foundation. 

\section*{Appendix}
\subsection*{The symmetry group of the model is a proper subgroup of the 
group of vertex translations} 

The symmetry group of the model (excluding $SO(3,1)$ symmetries) is 
the group 
of cell translations. It has dimension $4N_3$ and is a subgroup of the 
symmetry group of GLBF (the group of vertex translations) that has 
dimension $4N_0$. Thus, the fact that the symmetry group of the model is 
smaller than that of GLBF follows from the fact (to be shown bellow) that 
$N_0>N_3$ in locally Euclidean simplicial lattices. 

In three dimensions, the Euler number is zero, i.e., 
$N_0-N_1+N_2-N_3=\chi =0$ 
where $N_i$ is the number of $i$-dimensional simplices (points, links, 
faces, and cells). Then the difference between $N_0$ and $N_3$ is the same 
as that between $N_1$ and $N_2$. In a simplicial lattice 
there are three links in each face, and 
each link is shared by three or more faces (excluding lattices with zero 
volume cells). Thus, $N_1 \ge N_2$, the equality holding 
only in a case where 
every link is shared by three faces. A simplicial 
lattice with this connectivity can not 
be deformed in to a flat lattice as I explain now: 
It is easy to prove that in Euclidean three dimensional space one cannot 
draw a tetrahedron and its four neighbors in such a way that these five 
tetrahedra form a convex polyhedron. Start embedding one 
tetrahedron in $R^3$, 
then because every link is shared by three faces the four neighbors of the 
tetrahedron must have their faces identified with each other. Thus, the 
five tetrahedron embedded in $R^3$ form a convex polyhedron that has no 
boundary faces. The contradiction indicates that it is impossible 
to draw a three-dimensional 
simplicial lattice with more than five cells where three faces 
share each link or equivalently only simplicial lattices with $N_0 > N_3$ 
fit locally in three-dimensional Euclidean space.

\end{document}